\documentclass[10pt,journal,final]{IEEEtran}

\ifCLASSINFOpdf
\usepackage[pdftex]{graphicx}
\else
\usepackage[dvips]{graphicx}
\fi
\usepackage[caption=false,font=normalsize,labelfont=sf,textfont=sf]{subfig}
\usepackage{amsfonts}
\usepackage{bm}
\usepackage{graphicx}	% 插入图片所需引入的宏包；
\usepackage{multicol}
\usepackage{multirow}
\usepackage{tabulary}
\usepackage{tabularx}
\usepackage{stfloats}
\usepackage{mathrsfs}
\usepackage{latexsym}
\usepackage{epsfig}
\usepackage{epstopdf}
\usepackage{array}
\usepackage{amsmath}
\usepackage{amssymb}
\usepackage{color, soul}
\usepackage{amsthm}
\usepackage{enumerate}
\usepackage{algpseudocode}
\usepackage{algorithm}
\usepackage{bm}
\usepackage{balance}
\usepackage[noadjust]{cite}
\usepackage{float}
\usepackage{cite}
\usepackage{graphicx}
\usepackage{colortbl}
\usepackage{flushend}
\usepackage{graphics}
\usepackage{url}
\usepackage{lettrine}
\usepackage{cite}

\UseRawInputEncoding
\theoremstyle{plain} %default

\def\bal#1\eal{\begin{align}#1\end{align}}

\begin{document}
	
\title{Coarse-to-Fine Semantic Communication Systems for Text Transmission} 

\author{Mengli Tao,  Jiancun Fan, \emph{Senior Member, IEEE},  Jie Luo, and Huiqiang Xie, \emph{Member, IEEE}

% \vspace*{-0.1in}
\thanks{ This work was supported by the National Natural Science Foundation of China under Grants No. 62471381 and No. 62401227. (\textit{Corresponding author: Jiancun Fan.})}
\thanks{Mengli Tao, Jiancun Fan, and Jie Luo are with the School of Information and Communication Engineering, Xi'an Jiaotong University, Xi'an, China (e-mail: taoml@stu.xjtu.edu.cn, fanjc0114@gmail.com, luojie@stu.xjtu.edu.cn).}
\thanks{Huiqiang Xie is with the College of Information Science and Technology, Jinan University, Guangzhou, China (e-mail: huiqiangxie@jnu.edu.cn).}
}
	
\IEEEpubid{\begin{minipage}{\textwidth}\ \\[35pt] \centering
		Copyright \copyright 2025 IEEE. Personal use of this material is permitted. 
		However, permission to use this material for any other purposes must \\ be obtained 
		from the IEEE by sending an email to pubs-permissions@ieee.org.
\end{minipage}}

\maketitle

\begin{abstract}
Achieving more powerful semantic representations and semantic understanding is one of the key problems in improving the performance of semantic communication systems. This work focuses on enhancing the \textcolor{black}{semantic understanding} of the text data to improve the effectiveness of semantic exchange. We propose a novel semantic communication system for text transmission, in which the semantic understanding is enhanced by coarse-to-fine processing. Especially, a dual attention mechanism is proposed to capture both the coarse and fine semantic information. \textcolor{black}{Numerical experiments show the proposed system outperforms the benchmarks in terms of bilingual evaluation, sentence similarity, and robustness under various channel conditions.}
\end{abstract}


\section{Introduction}

Shannon and Weaver \cite{art2} categorized communication into three levels: i) the transmission of symbols; ii) the semantic exchange of transmitted symbols; and iii) the effect of semantic information exchange. In the past decades, conventional communications aimed to guarantee accurate transmission at the bit level, which refers to the first level. As the communication paradigm shifts from bit delivery (first level) to semantic delivery (second/third level), semantic communications \cite{art42} show its superiority in alleviating the scarcity of spectrum and improving system robustness, which has been considered one of the potential techniques in sixth-generation wireless systems to support intelligent services, e.g., industrial internet of things, holographic communication, and virtual reality.

Deep learning (DL) based semantic communications is capable of transmitting different modalities of data in an end-to-end manner with the tolerance of transmission errors, which has attracted the attention of academics \cite{rao2018variable, art34, art27} and industries \cite{art41, art42}. Meanwhile, text-based semantic communications have been developed a lot thanks to the development of DL-based natural language processing (NLP) techniques, which can be roughly divided into \textit{recurrent neural network (RNN)-based} and \textit{Transformer-based} semantic communication (SemCom) systems.

\textit{RNN-based} SemCom systems employ RNN neural networks to understand the text and extract the semantics behind the text. Farsad \textit{et al.} \cite{art34} have developed the initially RNN-based joint source-channel codes for text transmission, in which the bidirectional RNN models the sequential dependencies between words and phrases in both directions of the sequence. Rao \textit{et al.} \cite{rao2018variable} have extended \cite{art34} to the variable-length codes with the binarization step. Dam \textit{et al.} \cite{Dam10048944} have designed a long short-term memory (LSTM) based SemCom system for text summarization tasks, in which the LSTM can detain long-term dependencies in the document. However, RNNs face challenges in semantic understanding and parallel computation, which limits the potential for semantic understanding and representations. The self-attention-based Transformer \cite{art18} overcomes the inherent limitations of RNNs. Therefore, Transformer-based SemCom systems have attracted the attention.

\textit{Transformer-based} SemCom systems adopt the Transformer as the main component to extract the semantics. Xie \textit{et al.} \cite{art4} have designed the initially Transformer-based semantic communication system (DeepSC), which achieves a higher BLEU score than the separation codes in the low signal-to-noise ratio (SNR) region. Kadam \textit{et al.} \cite{Kadam10319829} have introduced the keywords extractor in the DeepSC with the help of shared knowledge, such that removes the redundancy in the sentence at the word level and reduces the number of transmitted symbols. Similarly, Yi \textit{et al.} \cite{yi2023deep} have developed a shared knowledge-based semantic communication system to remove the semantic redundancy at the sentence level, in which Transformer is employed to form the semantic extractor and integrator. Besides, Wang \textit{et al.} \cite{art7} have employed the graph Transformer to extract the knowledge graph (KG) of the sentence as the semantics, which can further reduce the number of transmitted symbols. Zhou \textit{et al.} \cite{Zhou10262128} have also viewed the KG as the semantics and transmitted it over the air, in which the pre-training model Text-to-Text Transfer Transformer model is employed to recover the text from the received KG. These works show their superiority and flexibility to RNN-based works. 

The power of Transformer-based SemCom is rooted in the self-attention mechanism, i.e., \textcolor{black}{the spatial attention mechanism interacts each word with all other words in the sentence. By calculating the similarity between each word, it captures sentence-level semantics, which introduces the strong ability of semantic understanding and allows more efficient semantic exchange. However, the spatial attention mechanism only extracts the coarse semantics and ignores the fine semantics, which leads to semantic differences during sentence generation}. This letter focuses on enhancing the \textcolor{black}{semantic understanding} of the text data to improve the effectiveness of semantic exchange. We propose a novel semantic communication system for text transmission, in which the semantic understanding is enhanced by coarse-to-fine processing. The main contributions can be summarized as follows,
\begin{itemize}
\item We propose a coarse-to-fine SemCom system for text transmission, in which both the coarse and fine semantics are extracted and fused to enable more powerful semantic understanding. 
\item The coarse-to-fine SemCom system consists of the dual attention-based Transformer, in which spatial and channel-wise attention mechanisms extract the coarse and fine semantics, respectively. 
\end{itemize}

The rest of the paper is organized as follows. In Section II, the system model is presented. In Section III, the coarse-to-fine SemCom system for text transmission is introduced, in which the spatial and channel-wise attention mechanisms are described in detail. Section IV presents the numerical results to show the performance of the proposed model. Finally, conclusions are given in Section V.

\textit{Notation}: Throughout this paper, bold letters are used to denote vectors or matrices, and regular letters are used to denote variables. $\mathbb{R}$ and $\mathbb{C}$ denotes the real and complex fields, respectively. ${( \cdot )^T}$ denotes the matrix transpose, and $w\sim {\cal CN}(\mu,\sigma _n^2)$ denotes that the variable $w$ follows the circularly complex Gaussian distribution with mean $\mu$ and covariance $\sigma _n^2$. ${\bm X}[k]$ represents the $k$-th row of the ${\bm X}$. ${\bm x}[k]$ represents the $k$-th value of the ${\bm x}$.

\section{System Model}
As shown in Fig. 1, we consider a single-input single-output (SISO) scenario. The system model consists of a transmitter, physical channel, and receiver. The transmitter includes the semantic encoder and channel encoder, where the transmitter and the receiver share the knowledge base (KB). The receiver consists of the channel decoder and semantic decoder to reconstruct the transmitted data. The transceiver shares the background knowledge so that the whole system learns the common knowledge of the particular target.

\begin{figure}[!t]
    \centering
    \includegraphics[width=9cm]{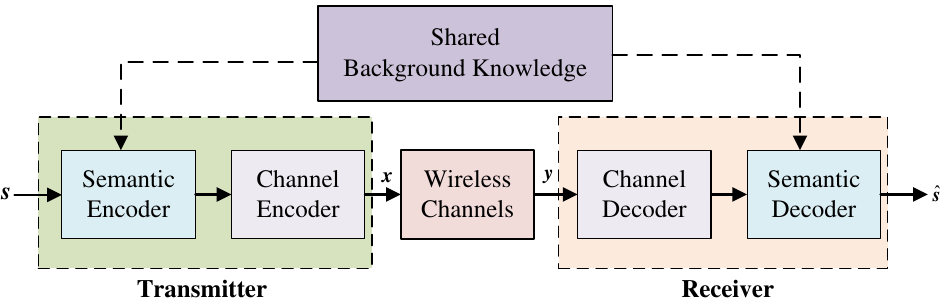}
    \captionsetup{justification=centering}
    \caption{The proposed semantic communication system model.}
    \label{Fig. 1.}
\end{figure}

Given a sentence with length $L$, $\bm{s}$, in which ${s[k]}$ is the $k$-th word in $\bm{s}$. Semantic encoding and channel encoding are performed to extract and transmit data, which is given by
\begin{equation}
    \bm{x} = {C_{\bm{\alpha}} }({S_{\bm{\beta}} }(\bm{s})),
\end{equation}
where ${\bm x} \in \mathbb{C}^{M\times 1}$,  $\bm{\beta} $-parameterized network ${S_{\bm{\beta}} }$ and ${\bm{\alpha}}$-parameterized net ${C_{\bm{\alpha}}}$ correspond to semantic encoder and channel encoder, respectively.

Then, the encoded symbols are transmitted over the block fading channel. The received signal is represented as 
\begin{equation}
 \bm{y} = {h}\bm{x} + \bm{n},
\end{equation}
where $\bm{y} \in {\mathbb{C}^{M \times 1}}$, ${h}$ is the channel gain parameter, and $\bm{n}$ denotes the additive white Gaussian noise (AWGN) term that satisfies $\bm{n}\sim {\cal CN}(0,\sigma _n^2 {\bf I})$.

At the receiver side, channel decoder and semantic decoder are employed to reconstruct the  sentences, which is given by
\begin{equation}
\bm{\hat s} = S_{\bm{\chi}} ^{ - 1}(C_{\bm{\delta}} ^{ - 1}(\bm{y})),
\end{equation}
where $\bm{\hat s}$ denotes the recovered sentence, and $\bm{\delta }$-parameterized net $C_{\bm{\delta}} ^{ - 1}$ and $\chi  $-parameterized net $S_{\bm{\chi}} ^{ - 1}$ correspond to channel decoder and semantic decoder, respectively.

\begin{figure*}[htbp]
\centering
\includegraphics[width=17cm]{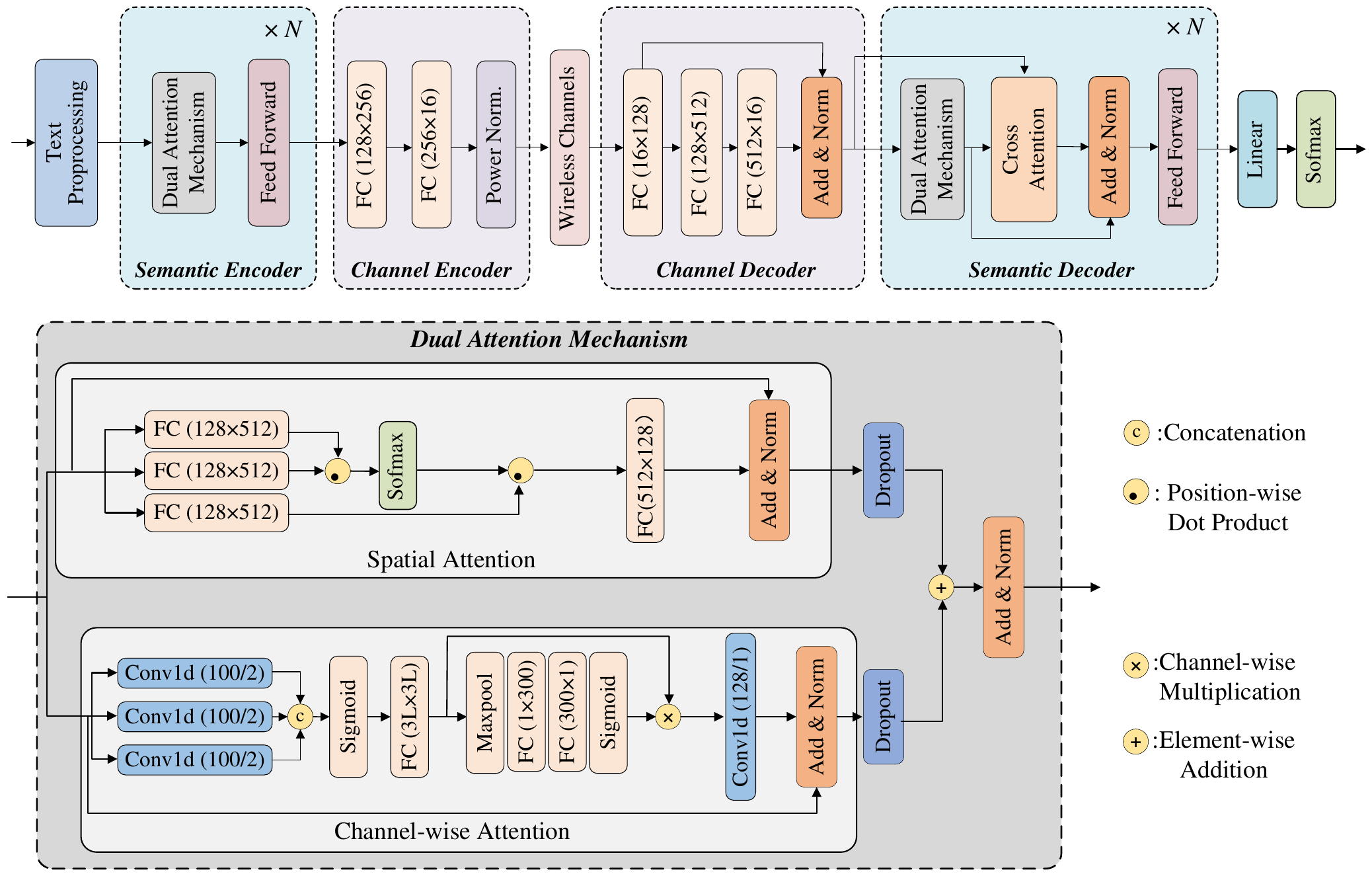}
\captionsetup{justification=centering}
\caption{\textcolor{black}{The overall neural network architecture of the proposed system.}}
\label{Fig. 2.}
\end{figure*}	

\section{The Proposed Coarse-to-Fine SemCom System}
In this section, we first introduce the designed system in detail. Then, we describe the coarse-to-fine processing to show the spatial and channel-wise attention mechanisms. Finally, the training details are given.

\subsection{The Model Description}
The proposed SemCom system is shown in Fig. 2. \textcolor{black}{The transmitter consists of a semantic encoder and a channel encoder. The semantic encoder is composed of several refined Transformer encoder layers with a dual attention mechanism, used to extract coarse and fine semantic features from the text to be transmitted, while the semantic decoder includes Transformer decoder layers based on the dual attention mechanism. The reasons behind the design can be summarized as follows. One is by integrating channel attention and spatial attention, the model can handle different types of features simultaneously, producing more comprehensive and accurate representation results. Another reason is that in order to be able to extract semantics more accurately, both kinds of attention will be applied directly to the original features, designed as a parallel structure rather than a serial one.}

\textcolor{black}{The channel encoder uses dense layers with different units to generate symbols for subsequent transmission. Accordingly, the receiver consists of a channel decoder for symbol detection and a semantic decoder for text estimation. The channel decoder uses dense layers with different units. The reasons behind the design can be summarized as follows. One reason is that using dense layers with units of different sizes helps improve the model's adaptability and generalization ability. Another reason is that it can simulate signal processing at different scales, thus improving the accuracy and robustness of encoding and decoding.}

The key behind the design is the dual attention mechanism, which enables coarse-to-fine semantic processing and semantic delivery with high fidelity.

\subsection{The Coarse-to-Fine Processing}
The coarse-to-fine processing mainly includes spatial attention and channel attention mechanisms. \textcolor{black}{The spatial attention mechanism extracts coarse-scale information. It first calculates the importance of each word in a sentence, then fuses the semantic information of different words based on the computed importance, and finally outputs semantic information at the sentence level. In parallel with the spatial attention mechanism, the channel attention mechanism captures relationships between phrases, words, or characters of different lengths by using convolution kernels of varying sizes. Therefore, the channel attention mechanism is able to understand fine semantics. Finally, a fusion block to fuse the coarse and fine semantics.}

\subsubsection{Spatial Attention Mechanism}

The spatial attention mechanism is the same as the self-attention mechanism in the original Transformer. Given the input of $i$-th layer in the dual-attention Transformer, ${\bm Z}_{i} \in \mathbb{R}^{ L \times d}$, in which ${\bm Z}_{i}[k]$ is the semantic feature of the ${\bm s}[k]$. Then, the spatial attention mechanism at the $i$-th layer is given by
\begin{equation}\label{eq-4}
 {f_{\rm SA}}({\bm Z}_i) = LN({\rm{softmax}}(\frac{{{{\bm Q}_i}{\bm K}_i^T}}{{\sqrt {{d}} }}){{\bm V}_i}{\bm W}^{F}_i + {\bm Z}_i),
\end{equation}
where \textcolor{black}{$LN(\cdot)$ denotes the layer normalization}, ${{\bm Q}_i}={\bm Z}_i {\bm W}^{Q}_i $, ${{\bm K}_i}= {\bm Z}_i {\bm W}^{K}_i$, and ${{\bm V}_i}= {\bm Z}_i {\bm W}^{V}_i$, in which ${\bm W}^{F}_i$, ${\bm W}^{Q}_i, {\bm W}^{K}_i$, and ${\bm W}^{V}_i$ are the learnable weights at the $i$-th layer. \text{softmax}($\cdot$) is the softmax activation.

From \eqref{eq-4}, we can observe that the spatial attention mechanism characterizes the relationship between words by ${\bm Q}_i{\bm K}^T_i$, i.e., the context. \textcolor{black}{This context indicates the similarity between a word and all other words in the sentence. The higher the value of a certain weight, the greater the similarity between the corresponding word vectors. Then, a new word vector is obtained by multiplying the $\bm{V}_i$, i.e., input word vectors, and the spatial attention matrix, which refers to the semantics at the sentence level. Then, we get the semantic features of the $i$-th word according to the context.} Therefore, the spatial attention mechanism is capable of understanding the sentence at the sentence level. However, the spatial attention mechanism focuses on the sentence level but ignores the word level, which means that the fine semantic information is ignored.

\subsubsection{Channel-wise Attention Mechanism}

The channel-wise attention mechanism learns the characters of each word by performing the attention mechanism at the word level. The channel-wise attention mechanism at the $i$-th layer consists of the following components.

First, the attention map is 
\begin{equation}\label{eq-5}
\textcolor{black}{
\begin{array}{l}
{f_{{\rm{CA}}}}({{\bm Z}_i}) = LN({f^C}({\rm{sigmoid(}}{\bm W}_{_2}^C{\bm W}_{_3}^C{\rm{(MP}}({{\tilde {\bm Z}}_i})))\\
\begin{array}{*{20}{c}}
{\begin{array}{*{20}{c}}
{\begin{array}{*{20}{c}}
{}&{}
\end{array}}&{}
\end{array}}&{}
\end{array} \odot {{\tilde {\bm Z}}_i})^T + {{\bm Z}_i})
\end{array},}
\end{equation}
where \textcolor{black}{$LN(\cdot)$ denotes the layer normalization}, \textcolor{black}{$f^{C}(\cdot)$ is the one dimension (1D)} convolutional layer, \textcolor{black}{${\tilde  {\bm Z}_i} = {\bm W}^C_1({\rm{sigmoid}}([{{\bm C}_1};{{\bm C}_2};{{\bm C}_3}]))$}, ${\bm C}_1=f^{C_1}_i({\bm Z}_i^T)$, ${\bm C}_2=f^{C_2}_i ({\bm Z}_i^T)$, and ${\bm C}_3=f^{C_3}_i ({\bm Z}_i^T)$, where $f^{C_1}_i(\cdot)$, $f^{C_2}_i(\cdot)$, and $f^{C_3}_i(\cdot)$ are the 1D convolutional layers with different kernel sizes, \text{MP}$(\cdot)$ is the maxpool operation. The 1D convolutional layer is performed along the semantic feature of each word, such that it captures the semantics at the word level. 

From \eqref{eq-5}, we can observe that the channel-wise attention mechanism uses ${{\bm C}_1},{{\bm C}_2},{{\bm C}_3}$ with convolution kernels of different sizes. This approach effectively operates based on the semantic features of each word, capturing word-level semantic information. \textcolor{black}{In this process, each channel typically corresponds to a specific feature map resulting from applying learnable filters to the input data, typically representing different learned patterns of features \cite{art29}. Therefore, the attention mechanism can be employed to process different types of features.} 

\subsubsection{Spatial-Channel Fusion Block}
To fuse these two parallel modules, we use a simple addition operation to fuse the features of these two attention blocks 
\begin{equation}
\textcolor{black}{
{\bm Z}_{i+1} = f_{\rm FFN}(f_{\rm SA}({\bm Z}_i) + f_{\rm CA}({\bm Z}_i)),}
\end{equation}
where $f_{\rm FFN}(\cdot)$ is the feed-forward function in the original Transformer.
			
\subsection{Loss Function}
To minimize the difference between $\bm{s}$ and $\bm{\hat s}$, we use the cross-entropy (CE) loss function,
\begin{equation}\label{loss-func}
{{\mathcal L}_{\rm CE}}(\bm{s}, \bm{\hat s}) =  - \mathbb{E}_{{\bm s} \sim p({\bm s}) } \left[\log (q({\bm {\hat s}})) \right],
\end{equation}
where $p({\bm s})$ is the true probability and $q({\bm {\hat s}})$ is the predicted probability of the predicted sentence $\bm{\hat s}$. CE can measure the difference between these two distributions. By minimizing the loss function, the semantic accuracy of text transmission is guaranteed.

\begin{table}
	\renewcommand{\arraystretch}{1.4}
	\begin{center}
		\caption{The Settings Of The Network.}
		\label{tab1}
		\begin{tabular}{| c | c | }	
			\hline
			& Proposed \\
			\hline
			\multirow{3}*{Transmitter} & 3$\times$Dual-Attention Transformer (8 heads) \\
			\cline{2-2}
			& Dense (256 units) \\
			\cline{2-2}
			& Dense (16 units) \\
			\hline
			Channel & AWGN/Fading channel  \\
			\hline
			\multirow{3}*{Receiver} & Dense (128 units) \\
			\cline{2-2}
			& Dense(512 units) \\
                \cline{2-2}
			& Dense(128 units) \\
			\cline{2-2}
			& 3$\times$Dual-Attention Transformer (8 heads) \\
			\cline{2-2}
			& Prediction Layer (Dictionary Size)\\
			\hline
		\end{tabular}
	\end{center}		
\end{table}
	
\section{Simulation Results}
In this section, simulation results versus BLEU scores \textcolor{black}{\cite{art35}} and similarity scores are provided to verify the effectiveness of the proposed model. 
	
\subsection{Simulation Settings}
\subsubsection{Dataset}
The English version of the proceedings of the European Parliament dataset is used as the dataset, in which, unless stated otherwise, sentences are 4-36 words and divided into training and test parts.

\subsubsection{Implementation Details}
We present the structure of the proposed system in Table 1. We set up three Transformer encoder and decoder layers with 8 heads, and the channel encoder and decoder are set up as dense layers with 128, 512 and 128 units, respectively. It is assumed that perfect channel state channel state information (CSI) is obtained for all schemes.
\textcolor{black}{
\subsubsection{Training parameters setting}
In our experiments, unless otherwise specified, we set up three Transformer encoder and decoder layers with 8 heads. The batch size and number of epoch are set to 256 and 80, respectively. We used the Adam optimizer with a learning rate of 0.0005 and employed \eqref{loss-func} as the loss function to improve training accuracy.
}

\subsubsection{Benchmarks}
For the baseline, we compare the proposed system with the benchmarks as follows: 
\begin{itemize} 
    \item DeepSC \cite{art4}  consists of three Transformer Layers
    \item RNN-based joint source-channel coding \cite{art39} consists of two-way long short-term memory (BLSTM) layers
    \item Separate source and channel coding utilize Huffman coding and Turbo coding\cite{art31} as the source coding and channel coding, respectively.
\end{itemize}

\subsection{Performance comparison}
\begin{figure}[!t]
    \begin{minipage}{1.04\linewidth}
        \centering
        \includegraphics[width=85mm]{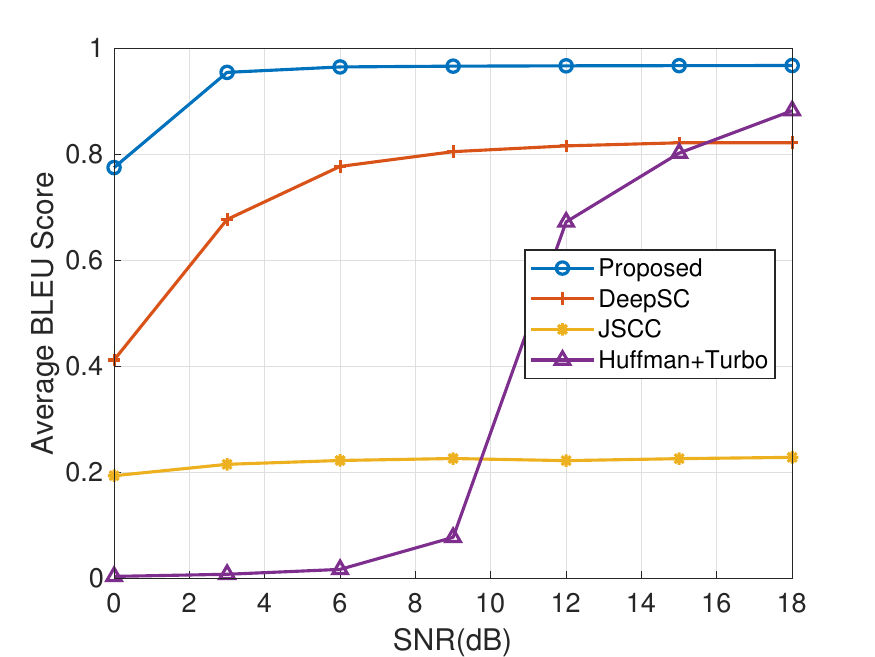}
        \centerline{(a) AWGN channels.}
    \end{minipage}

    \begin{minipage}{1.04\linewidth}
        \centering
        \includegraphics[width=85mm]{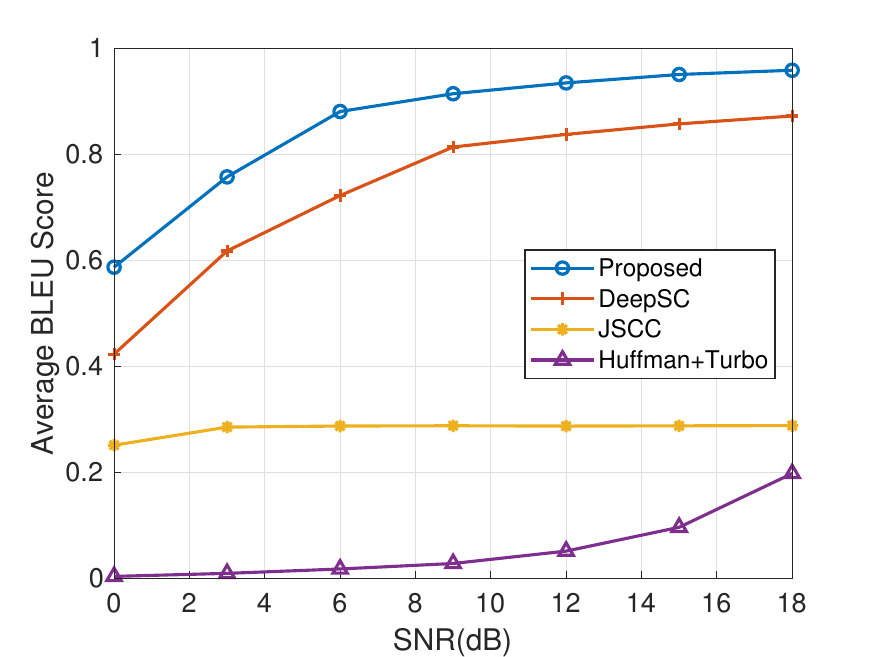}
        \centerline{(b) Rayleigh fading channels.}
    \end{minipage}%
\centering
\caption{BLEU score versus SNR over the Rayleigh fading channel and the AWGN channels.}
\end{figure}

Fig. 3 shows \textcolor{black}{the average BLEU score for 1 to 4 gram} versus SNR over different channels. For AWGN channels, Fig. 3(a) shows that the proposed model outperforms all other schemes in terms of average BLEU score in the entire SNR regime. In particular, the proposed system achieves a higher average BLEU score than DeepSC, which proves the effectiveness of the coarse-to-fine semantic fusion. When SNR is greater than 15 dB, the average BLEU score of Huffman with Turbo coding approaches that of the proposed system due to the accurate transmission characteristics of conventional communications. \textcolor{black}{In high SNR regions, the error correction offered by channel coding enables the "Huffman+Turbo" scheme to outperform the DeepSC scheme. Another reason is that DL-based methods belong to a kind of lossy compression, which compresses the dimensionality of semantic representations and loses a little information.} For Rayleigh channels, Fig. 3(b) shows that the proposed system experiences slight performance degradation in terms of average BLEU score and still outperforms all the benchmarks. However, the conventional schemes have a large performance gap compared with those in Fig. 3(a). This indicates the high robustness of the proposed system.

\begin{table*}
    \centering
    \renewcommand{\arraystretch}{1.25}
    \caption{The Decoded Sentences By Different Methods Over Rayleigh Fading Channels When SNR Is 18DB}
    \begin{tabular}{|c|l|}
        \hline
        Method   & \multicolumn{1}{c|}{Sentence}  \\ \hline
        Transmitted sentence   & \begin{tabular}[c]{@{}l@{}}last year we introduced activity based budgeting and this year we are creating a clear link between  strategic \\objectives and the allocation of resources especially human resources .\end{tabular} \\ \hline
        Proposed & \begin{tabular}[c]{@{}l@{}}last year we introduced activity based budgeting and this year we are creating a clear link between strategic \\objectives and the allocation of resources especially human resources .\end{tabular}  \\ \hline
        DeepSC   & \begin{tabular}[c]{@{}l@{}}the we border activity based . and this year we are in a clear link between strategic objectives and the  allocation \\ of resources especially human resources.\end{tabular}                            \\ \hline
        JSCC    & \begin{tabular}[c]{@{}l@{}}sevolle we bureaucratic kinnock ukraine ? and taxpayers usd . veto i enlargement iraq un finland stockholm\end{tabular}  \\ \hline
        Huffman + Turbo   & \begin{tabular}[c]{@{}l@{}}last yarl er indeth acti  sste based budget fio dor himk we sdpwoed djfhcnd an coplding laarnd . bwill  shniks \\skpsf already ? of tranrneu. .\end{tabular} \\ \hline
    \end{tabular}
    \label{tab-2}
\end{table*}

\begin{figure}[!t]
    \begin{minipage}{1.04\linewidth}
        \centering
        \includegraphics[width=85mm]{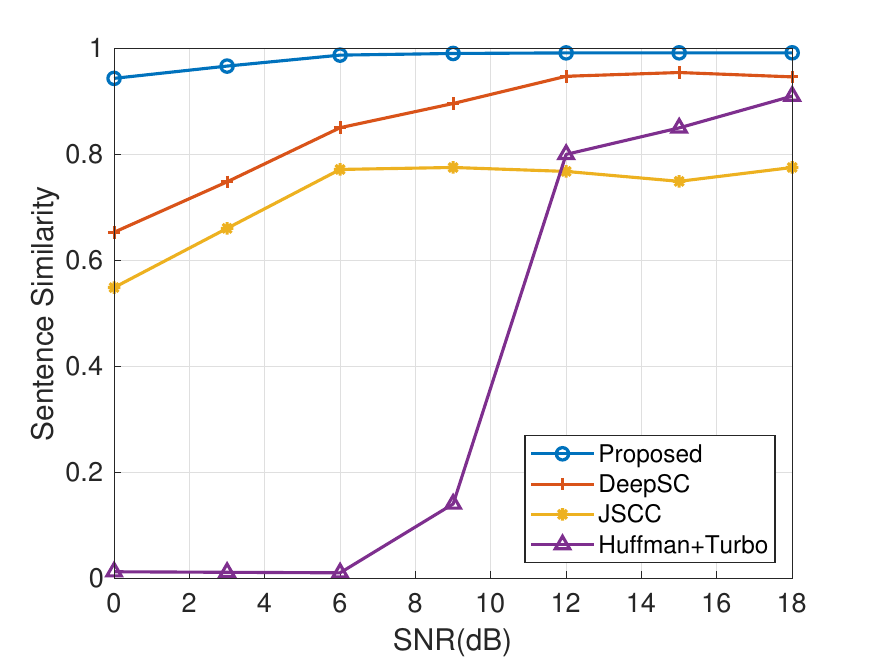}
        \centerline{(a) AWGN channels.}
    \end{minipage}

    \begin{minipage}{1.04\linewidth}
        \centering
        \includegraphics[width=85mm]{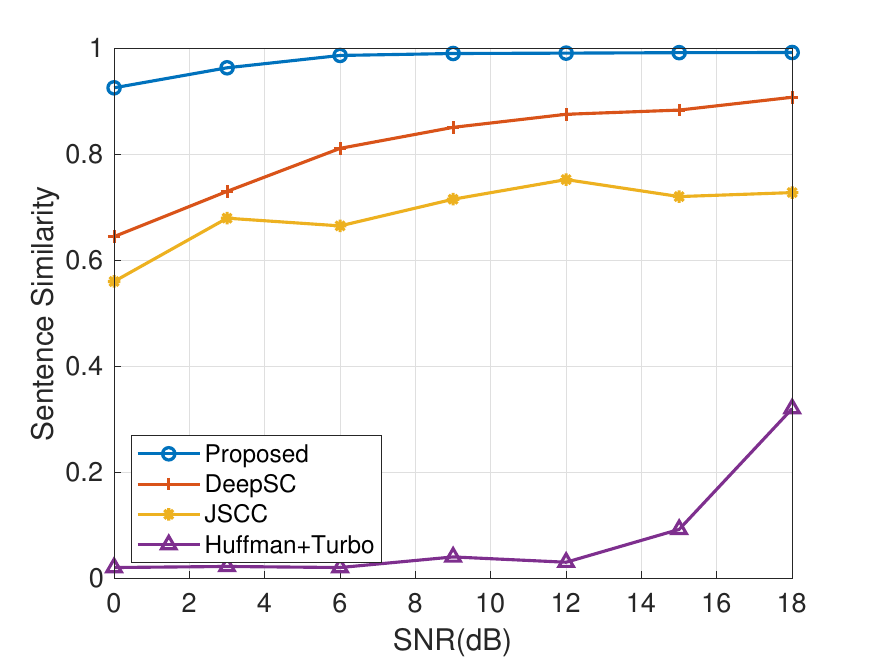}
        \centerline{(b) Rayleigh fading channels.}
    \end{minipage}%
\centering
\caption{Similarity score versus SNR over AWGN and Rayleigh fading channels.}
\end{figure}
    
\begin{table}
    \renewcommand{\arraystretch}{1.4}
    \begin{center}
    \caption{\textcolor{black}{The Scores By Different Methods Over Rayleigh fading Channels When SNR Is \textcolor{black}{18DB} And 0DB}}
    \begin{tabular}{|c|c|c|}
        \hline
        & \multicolumn{2}{c|}{SNR = 18 dB} \\  % Add SNR label over both columns
        \hline
           & BLEU Score & Sentence Similarity Score \\
        \hline
        Proposed(4-36 words)   & 0.92 & 0.99 \\ 
        \hline
        Proposed(37-100 words) & 0.92 & 0.99 \\ 
        \hline
        DeepSC(4-36 words)     & 0.81 & 0.89 \\ 
        \hline
        DeepSC(37-100 words)  & 0.65 & 0.69 \\ 
        \hline
        & \multicolumn{2}{c|}{SNR = 0 dB} \\  % Add SNR label over both columns
        \hline
           & BLEU Score & Sentence Similarity Score \\
        \hline
        Proposed(4-36 words)   & 0.79 & 0.95 \\ 
        \hline
        Proposed(37-100 words) & 0.72 & 0.88 \\ 
        \hline
        DeepSC(4-36 words)     & 0.6 & 0.65 \\ 
        \hline
        DeepSC(37-100 words)  & 0.43 & 0.53 \\ 
        \hline
    \end{tabular}
    \label{tab-3}
    \end{center}
\end{table}

Fig. 4 shows the sentence similarity scores versus SNR over different channels. The similarity score is computed by the BERT \cite{art4}. In Fig. 4(a), the sentence similarity of the proposed system also outperforms all the benchmarks in the whole SNR region. The semantic similarity scores are improved compared with the BLEU scores in Fig. 3, which proves that the semantic communication methods are more advantageous in recovering the meanings of the text. Change from AWGN to Rayleigh fading channels, Fig. 4(b) shows that the proposed system experiences slight performance degradation in terms of similarity score and still outperforms all the benchmarks. In particular, semantic similarity of the proposed system drops smaller value than the DeepSC when channels are changed from AWGN to Rayleigh fading. This indicates that the proposed system has better ability of protecting meanings over different channels. The visualization results are shown in Table II. \textcolor{black}{In Fig. 4, we can also observe that the sentence similarity of semantic communication methods outperforms conventional methods. Additionally, for the JSCC scheme, the sentence similarity performance shows slight fluctuations as the SNR increases. This is because of the limited model capacity. During the training with physical channels, the model has to focus on learning to deal with noise and allocates less attention to extracting semantics. Therefore, the model cannot extract the semantics accurately, which causes unstable sentence generation.}

To compare the ability to capture contextual information over Rayleigh fading channels, we selected sentences with over 36 words, but no more than 100 words, which increases the difficulty of sentence transmission to some extent. \textcolor{black}{Among them, all the parameters are set the same as above, with only the sentence length changed. Table \ref{tab-3} shows the sentence similarity and BLEU 1-gram scores when SNR = 18 dB and SNR = 0 dB. It can be observed that the proposed system consistently outperforms DeepSC as sentence length significantly increases. Besides, the scores of the proposed system show less variation compared to the DeepSC approach with changes in SNR, indicating that the proposed system has an advantage in transmitting longer sentences.}

\section{Conclusion}
In this work, a dual attention-based semantic communication system is proposed for text transmission. We use two parallel attention mechanisms for semantic coders, where spatial attention is more concerned with coarse semantics of text, and the fine semantics of text are also crucial for text understanding, to fully exploit the semantic information, a channel-wise attention mechanism is designed to capture fine semantics of text. Then, the coarse and fine semantics are fused to obtain a better semantic representation. Numerical comparisons show that the method proposed in this paper improves the BLEU score, the accuracy of decoding text semantics, and the similarity score in the presented SNR regime.

\bibliographystyle{IEEEtran}
\bibliography{reference1}

\begin{thebibliography}{10}
\providecommand{\url}[1]{#1}
\csname url@samestyle\endcsname
\providecommand{\newblock}{\relax}
\providecommand{\bibinfo}[2]{#2}
\providecommand{\BIBentrySTDinterwordspacing}{\spaceskip=0pt\relax}
\providecommand{\BIBentryALTinterwordstretchfactor}{4}
\providecommand{\BIBentryALTinterwordspacing}{\spaceskip=\fontdimen2\font plus
\BIBentryALTinterwordstretchfactor\fontdimen3\font minus
  \fontdimen4\font\relax}
\providecommand{\BIBforeignlanguage}[2]{{%
\expandafter\ifx\csname l@#1\endcsname\relax
\typeout{** WARNING: IEEEtran.bst: No hyphenation pattern has been}%
\typeout{** loaded for the language `#1'. Using the pattern for}%
\typeout{** the default language instead.}%
\else
\language=\csname l@#1\endcsname
\fi
#2}}
\providecommand{\BIBdecl}{\relax}
\BIBdecl

\bibitem{art2}
C.~E. Shannon and W.~Weaver, ``The mathematical theory of communication,''
  \emph{The University of Illinois Press}, vol.~29, no.~1, pp. 69--93, Mar.
  1953.

\bibitem{art42}
Z.~Qin, X.~Tao, J.~Lu, W.~Tong, and G.~Y. Li, ``Semantic communications:
  principles and challenges,'' \emph{arXiv preprint arXiv:2201.01389}, Jun.
  2022.

\bibitem{rao2018variable}
M.~Rao, N.~Farsad, and A.~Goldsmith, ``Variable length joint source-channel
  coding of text using deep neural networks,'' in \emph{2018 IEEE 19th
  international workshop on signal processing advances in wireless
  communications (SPAWC)}.\hskip 1em plus 0.5em minus 0.4em\relax IEEE, 2018,
  pp. 1--5.

\bibitem{art34}
N.~Farsad, M.~Rao, and A.~Goldsmith, ``Deep learning for joint source-channel
  coding of text,'' in \emph{in Proc. IEEE Int. Conf. Acoust., Speech and
  Signal Process. (ICASSP)}, Apr. 2018, pp. 2326--2330.

\bibitem{art27}
H.~Xie, Z.~Qin, X.~Tao, and K.~B. Letaief, ``Task-oriented multi-user semantic
  communications,'' \emph{IEEE J. Sel. Areas Commun.}, vol.~40, no.~9, pp.
  2584--2597, Sep. 2022.

\bibitem{art41}
Q.~Lan, D.~Wen, Z.~Zhang, Q.~Zeng, X.~Chen, P.~Popovski, and K.~Huang, ``What
  is semantic communication? a view on conveying meaning in the era of machine
  intelligence,'' \emph{Journal of Communications and Information Networks},
  vol.~6, no.~4, pp. 336--371, Dec. 2021.

\bibitem{Dam10048944}
S.~K. Dam, M.~Shirajum~Munir, A.~D. Raha, A.~Adhikary, S.-B. Park, and C.~S.
  Hong, ``Rnn-based text summarization for communication cost reduction: Toward
  a semantic communication,'' in \emph{2023 International Conference on
  Information Networking (ICOIN)}, 2023, pp. 423--426.

\bibitem{art18}
A.~Vaswani, N.~Shazeer, N.~Parmar, J.~Uszkoreit, L.~Jones, A.~N. Gomez, L.~u.
  Kaiser, and I.~Polosukhin, ``Attention is all you need,'' in \emph{Proc. Adv.
  Neural Inf. Process. Syst.}, vol.~30.\hskip 1em plus 0.5em minus 0.4em\relax
  Long Beach, CA, USA., Dec. 2017, pp. 5998--6008.

\bibitem{art4}
H.~Xie, Z.~Qin, G.~Y. Li, and B.-H. Juang, ``Deep learning enabled semantic
  communication systems,'' \emph{IEEE Trans. Signal Process.}, vol.~69, pp.
  2663--2675, Apr. 2021.

\bibitem{Kadam10319829}
S.~Kadam and D.~I. Kim, ``Knowledge-aware semantic communication system design
  and data allocation,'' \emph{IEEE Transactions on Vehicular Technology},
  vol.~73, no.~4, pp. 5755--5769, Apr. 2024.

\bibitem{yi2023deep}
P.~Yi, Y.~Cao, X.~Kang, and Y.-C. Liang, ``Deep learning-empowered semantic
  communication systems with a shared knowledge base,'' \emph{IEEE Transactions
  on Wireless Communications}, Early Access, 2023.

\bibitem{art7}
Y.~Wang, M.~Chen, W.~Saad, T.~Luo, S.~Cui, and H.~V. Poor, ``Performance
  optimization for semantic communications: An attention-based learning
  approach,'' \emph{IEEE J. Sel. Areas Commun.}, vol.~40, no.~9, pp.
  2598--2613, Sep. 2022.

\bibitem{Zhou10262128}
F.~Zhou, Y.~Li, M.~Xu, L.~Yuan, Q.~Wu, R.~Q. Hu, and N.~Al-Dhahir, ``Cognitive
  semantic communication systems driven by knowledge graph: Principle,
  implementation, and performance evaluation,'' \emph{IEEE Transactions on
  Communications}, vol.~72, no.~1, pp. 193--208, Jan. 2024.

\bibitem{art29}
J.~Hu, L.~Shen, and G.~Sun, ``Squeeze-and-excitation networks,'' in \emph{Proc.
  Conf. Comput. Vis. Pattern Recognit. (CVPR)}, Jun. 2018, pp. 7132--7141.

\bibitem{art35}
K.~Papineni, S.~Roukos, T.~Ward, and W.-J. Zhu, ``Bleu: a method for automatic
  evaluation of machine translation,'' in \emph{Proc. 40th Annu. Meeting Assoc.
  Comput. Linguistics}, Jul. 2002, pp. 311--318.

\bibitem{art39}
C.~Heegard and S.~B. Wicker, \emph{Turbo coding}.\hskip 1em plus 0.5em minus
  0.4em\relax Springer Science \& Business Media, 2013, vol. 476.

\bibitem{art31}
S.~Park, O.~Simeone, and J.~Kang, ``End-to-end fast training of communication
  links without a channel model via online meta-learning,'' in \emph{in Proc.
  21st IEEE Int. Workshop Signal Process. Adv. Wireless Commun.(SPAWC)}, Mar.
  2020, pp. 1--5.

\end{thebibliography}

\end{document}